\RequirePackage[hyphens]{url}
\documentclass[final,5p,times,twocolumn]{elsarticle}
\usepackage{lineno,hyperref}
\usepackage{amsmath}
\usepackage{siunitx}

\usepackage{array}
\usepackage{booktabs}

\usepackage[english]{babel}
\usepackage[autostyle]{csquotes}
\usepackage{float}
\usepackage{blindtext}
\usepackage{mathptmx}
\usepackage{graphicx}
\usepackage{subfig}
\usepackage{tabularx}
\usepackage{fleqn}
\usepackage{natbib}
\usepackage{newcent}  
\usepackage{amssymb}  
\usepackage{amsthm}   
\usepackage{epigraph} 
\usepackage{booktabs} 
\usepackage{fancyhdr} 
\usepackage{color}    
\usepackage{pdfpages} 
\usepackage{texshade} 
\usepackage[usenames,dvipsnames]{xcolor}
\usepackage[titletoc,title]{appendix}
\usepackage[dotinlabels]{titletoc}
\usepackage{lscape}
\usepackage{pdflscape}
\usepackage{gensymb}
\usepackage{rotating} 
\usepackage{textcomp}
\usepackage{float}
\usepackage{hyperref}
\usepackage[hyphenbreaks]{breakurl}

\usepackage[hyphens]{url}
\usepackage{breakcites}

 \biboptions{super,sort&compress }

\journal{Computational Materials Science}

\begin{document}

\begin{frontmatter}
\title{\textit{Combined Machine Learning and CALPHAD Approach for Discovering Processing-Structure Relationships in Soft Magnetic Alloys} }

\author[label1]{Rajesh Jha\corref{mycorrespondingauthor}}
\ead{rajeshjha@mines.edu}
\author[label2]{Nirupam Chakraborti}
\author[label3]{David R. Diercks}
\author[label1]{Aaron P. Stebner}
\author[label1]{Cristian V. Ciobanu\corref{mycorrespondingauthor}}
\ead{cciobanu@mines.edu}
\cortext[mycorrespondingauthor]{Corresponding author}

\address[label1]{Department of Mechanical Engineering, Colorado School of Mines, Golden, Colorado 80401, USA}
\address[label2]{Department of Metallurgical and Materials Engineering, Indian Institute of Technology Kharagpur, West Bengal, India}
\address[label3]{Department of Metallurgical and Materials Engineering, Colorado School of Mines, Golden, Colorado 80401, USA}

\begin{abstract}
FINEMET alloys have desirable soft magnetic properties due to the presence of
Fe$_3$Si nanocrystals with specific size and volume fraction. To guide future design of these alloys, we
investigate relationships between select processing parameters (composition, temperature, annealing time) and
structural parameters (mean radius and volume fraction)
of the Fe$_3$Si domains.
We present a combined CALPHAD and machine learning approach leading to well-calibrated metamodels able to predict structural parameters quickly and accurately for any desired inputs. To generate data, we have used a known precipitation model to
perform annealing simulations at a several temperatures, for
varying Fe and Si concentrations. Thereafter, we used the data to develop metamodels
for mean radius and volume fraction via the \emph{k}-Nearest Neighbour algorithm.
The metamodels reproduce closely the results from the precipitation  model over the entire annealing timescale. Our analysis via parallel coordinate charts shows the effect of composition, temperature, and annealing time, and helps
identify combinations thereof that lead to the desired mean radius and volume fraction for nanocrystals.
This work contributes to understanding the linkages between processing parameters and
desired microstructural characteristics responsible for achieving targeted properties, and illustrates ways to reduce the time from alloy discovery to deployment.
\end{abstract}

\begin{keyword}
\texttt{Soft magnetic alloys \sep FINEMET \sep CALPHAD \sep Machine Learning \sep k-Nearest Neighbour Algorithm \sep Parallel Coordinate Chart }
\end{keyword}

\end{frontmatter}


\section{INTRODUCTION}\label{introduction}

FINEMET alloys belong to a class of soft magnetic alloys based on the Fe-Si-Nb-B-Cu system.\cite{yoshizawa1988new}
In comparison with other soft magnets, FINEMET alloys possess high
saturation magnetization\cite{yoshizawa1988new} and
high permeability,\cite{willard2013nanocrystalline, App_Herzer, dia_hono1991atom, dia_lashgari2014composition} low core loss,\cite{yoshizawa1988new, willard2013nanocrystalline, App_Herzer, dia_lashgari2014composition} low magnetostriction,\cite{yoshizawa1988new, willard2013nanocrystalline, App_Herzer, dia_lashgari2014composition, dia_mattern1995effect} excellent temperature characteristics, small aging effects, and excellent high frequency characteristics.\cite{yoshizawa1988new, willard2013nanocrystalline, App_Herzer, dia_lashgari2014composition} As a result, FINEMET alloys have been successfully used in a number of applications including  choke coils,\cite{yoshizawa1988new, willard2013nanocrystalline, Dia_FINEMET_App, dia_herzer1993nanocrystalline, dia_herzer2013modern} mobile phones,\cite{willard2013nanocrystalline} noise reduction devices,\cite{willard2013nanocrystalline} computer hard disks,\cite{willard2013nanocrystalline} and transformers.\cite{yoshizawa1988new, willard2013nanocrystalline, dia_herzer1993nanocrystalline, dia_herzer2013modern, App_Herzer}  Superior soft magnetic properties are attributed to the nanocrystalline $\alphaα^{''}$-(Fe, Si) phase (Fe$_3$Si with D03 structure) in the size range of 10$-$15 nm diameter (radius 5$-$7.5 nm) and 0.7 volume fraction.\cite{yoshizawa1988new, willard2013nanocrystalline, dia_ayers1997model, dia_herzer1993nanocrystalline, dia_van1993nb, dia_herzer1991magnetization, dia_herzer2013modern, dia_herzer2010effect, App_Herzer, dia_hono1991atom, dia_lashgari2014composition, dia_clavaguera2002crystallisation, dia_conde1994crystallization, dia_mattern1995effect, dia_hono1999cu} Since its discovery, researchers have investigated FINEMET alloys to improve upon multiple soft magnetic properties by performing experiments followed by characterization using advanced diagnostic tools.\cite{willard2013nanocrystalline, dia_ayers1997model, dia_herzer1993nanocrystalline, dia_van1993nb, dia_herzer1991magnetization, dia_hono1991atom, dia_lashgari2014composition, dia_clavaguera2002crystallisation, dia_conde1994crystallization, dia_hono1999cu}

In materials design, understanding the various processing-structure-property (PSP) linkages plays an important role in designing advanced materials. In particular, correlations between microstructure and desired properties,\cite{1_PSP_Kalidindi, 2_PSP_Kalidindi, 3_PSP_Kalidindi, PSP_APL}  are essential for the
deployment of new materials into service. In addition,  composition variations and processing parameters (e.g., heat treatment schedule) play an integral role in modeling the microstructure(s) responsible for achieving  desired properties, where optimizing processing parameters along with composition  remains a challenging task.\cite{PSP_Challenges_Kalidindi} As an alternative to costly experimentation, the CALPHAD approach allows for investigating the effect of composition variations and heat treatment on the size distribution and volume fraction of the  phase(s) that are responsible for optimal or desired properties; indeed, it has been used for
studying soft magnets containing amorphous phases\cite{takeuchi2014thermodynamic, takeuchi2015thermodynamic, takahashi2017fe} using the commercial software Thermocalc.\cite{THERMOCALC} Recent studies indicate that simulations based on CALPHAD\cite{larsson2015scheme, pillai2016methods_CALPHAD} are in need of efficiency improvements if they are to be used for optimization of the composition and heat treatment schedule. To address this challenge, it is important to develop models that can both replicate maximum information available from prior studies and, in addition, demonstrate effectiveness in optimizing the processing protocol. This effectiveness should not come, for example, from repeating the same calculations at different compositions, but rather from learning the results obtained in several selected cases in order to predict the behaviour at other compositions.

Machine learning approaches  have been previous used to help reduce the time required in the alloy design process.\cite{jha2017SOM, jha2017magnetic_Book_Chapter, jha2016algorithms_JALCOM, jha2016combined_PhD_Thesis, fan_Jha_2016formation_IEEE, jha2015COBEM, jha2015magnetic_ACEX, jha2015algorithms_ICMM4, fan_jha2016evolution_JMMM, jha2015superalloy_MAMP, jha2014superalloy_ASME} Supervised machine learning approaches such as artificial neural networks,\cite{ann_pettersson2007genetic, jha2015superalloy_MAMP, jha2014_SRI, jha2014superalloy_ASME} \emph{k}-Nearest Neighbour algorithm ($k$-NN),\cite{K_NN_Algorithm, jha2014superalloy_ASME} genetic programming,\cite{biogp_giri2013genetic, jha2015superalloy_MAMP, jha2014_SRI, jha2014superalloy_ASME} kriging,\cite{kriging_pantula2017kernel, kriging_chugh2017data} and unsupervised approaches such as Principal Component Analysis (PCA),\cite{jha2016algorithms_JALCOM, jha2016combined_PhD_Thesis,jha2015algorithms_ICMM4} Hierarchical Clustering Analysis (HCA),\cite{jha2017magnetic_Book_Chapter, jha2016combined_PhD_Thesis, jha2015magnetic_ACEX} and Self Organizing Maps (SOM)\cite{jha2017SOM} have been previously used in materials science and can also be helpful in this case. From an implementation point of view, there exist several open-source software packages to develop response surfaces or metamodels using several different concepts from artificial intelligence. A machine learning model based on results from the CALPHAD approach  will serve as an important rapid screening tool before performing experiments and also in predicting outcomes in case of uncertainties in the composition of the material or in furnace temperature  during annealing.

In this article, we present a combined CALPHAD-machine learning approach for optimizing composition along with processing parameters
for FINEMET alloys by developing metamodels (response surfaces, or surrogate models) for the simulated crystallization of
Fe$_3$Si domains.
We have acquired data for mean radius and volume fraction of Fe$_3$Si nanocrystals through a recently developed
precipitation model\cite{jha2017FINEMET_CALPHAD} in Thermocalc,\cite{THERMOCALC} capable
of simulating the nucleation and growth of Fe$_3$Si nanocrystals from an amorphous phase.
Thereafter, we have used a \emph{k}-Nearest
Neighbour ($k$-NN) algorithm to generate computationally inexpensive metamodels to replace exhaustive Thermocalc modeling without any significant loss of accuracy. This way, we are able to demonstrate the efficacy of our combined CALPHAD-machine learning approach by predicting compositions and processing parameters that would lead to achieving the desired mean radius and volume fraction of Fe$_3$Si nanocrystals.
The developed metamodels capture the established nucleation and growth evolution\cite{kampmann1984decomposition, wagner1991homogeneous} within the CALPHAD approach, for the entire annealing timescale even for compositions and parameters that were not included in the training set for the metamodels. Another important observation is that the metamodels can predict outcomes in a fraction of the time taken by simulations in Thermocalc.\cite{THERMOCALC_TCFE8} Lastly, we propose Parallel Coordinates Charts (PCC)\cite{PCC}  for comprehensive visualization of the relationships between processing parameters and optimized quantities, and for  rapid identification of the parameters that lead to crystallization of Fe$_3$Si nanocrystals in the desired size  range and volume fraction. Our proposed approach helps reduce the alloy development time since it can serve as a tool for rapid screening of the multi-dimensional parameter space before performing experiments. As such, this combined machine learning and CALPHAD approach illustrates a case of addressing the challenge of simultaneously determining the effect of composition variation and processing parameters\cite{1_PSP_Kalidindi, 2_PSP_Kalidindi, 3_PSP_Kalidindi, PSP_APL, PSP_Challenges_Kalidindi} on the microstructure and can be extended to other alloy systems.

\section{METHODS}\label{methods}
\begin{figure}[ht]
\centering
\includegraphics[width = 6cm]{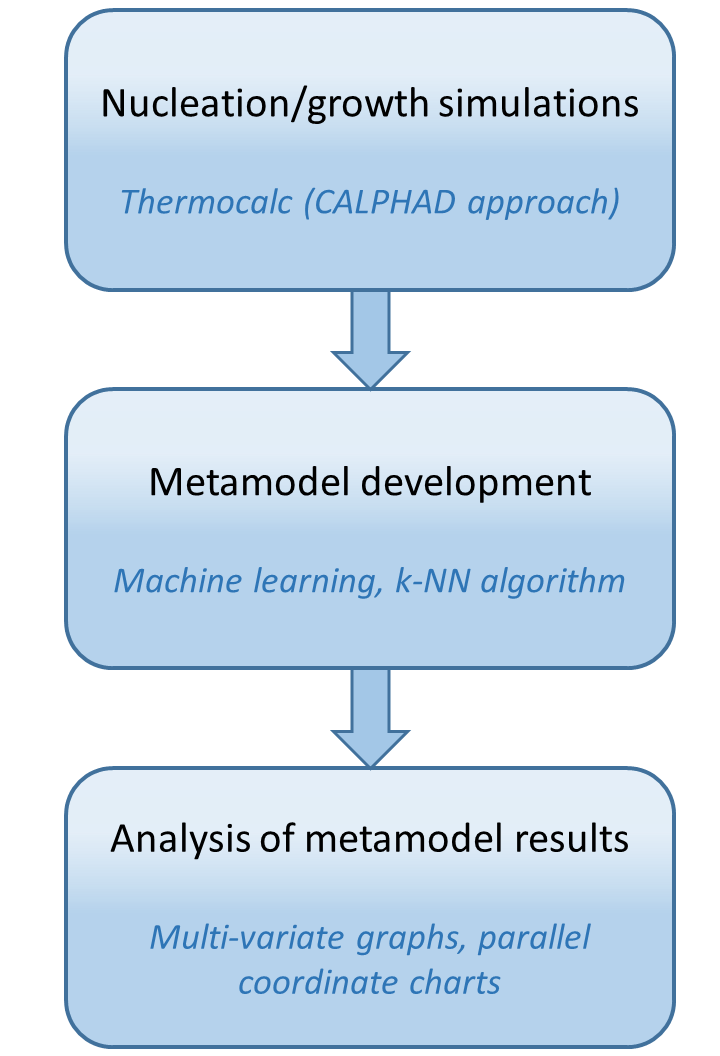}
\caption{Flowchart of steps followed in this work for a class of FINEMET alloy with composition Fe$_{72.89}$Si$_{16.21}$B$_{6.9}$Nb$_{3}$Cu$_{1}$}
\label{Paper_2_Flowchart}
\end{figure}

Figure~\ref{Paper_2_Flowchart} shows the schematic flowchart of the process we followed in order to develop our combined CALPHAD-machine learning approach for optimization
of nanocrystal size and volume fraction.  This approach is enabled by a nucleation and growth model (precipitation model) in Thermocalc,\cite{THERMOCALC_TCPRISMA} recently parameterized for FINEMET.\cite{jha2017FINEMET_CALPHAD} We used this model to generate data for mean radius and volume fraction of Fe$_3$Si nanocrystals grown upon annealing the
amorphous material, data which serves as a training set for  developing metamodels. Analysis of the results created by the metamodels reveals correlations between the input
parameters (composition, temperature, and time) and the optimized quantities. The three aspects (Figure~\ref{Paper_2_Flowchart}) are described in some detail below.

\subsection{Generating data for developing a metamodel}

To generate mean radius and volume fraction data, \cite{CITRINE} we have used the
TC-PRISMA\cite{THERMOCALC_TCPRISMA} module in Thermocalc, which relies on thermodynamic (TCFE8)\cite{THERMOCALC_TCFE8}  and mobility\cite{THERMOCALC_TCFE8_MOBFE3} databases.
TC-PRISMA\cite{THERMOCALC_TCPRISMA} uses the Kampmann-Wagner Numerical (KWN) method\cite{kampmann1984decomposition, wagner1991homogeneous} for simulating
nucleation and growth of precipitates during annealing. The KWN method is an extension of the Langer-Schwartz approach\cite{langer1980kinetics} and its modified form.\cite{COMPUTHERM}
To use the precipitation model, several input quantities in TC-PRISMA\cite{THERMOCALC_TCPRISMA} were previously parameterized\cite{jha2017FINEMET_CALPHAD} so that the
precipitation model simulates specifically and accurately the nucleation and growth of Fe$_3$Si nanocrystals  during
annealing.
The FINEMET base composition is  Fe$_{82.35}$Si$_{9.21}$B$_{1.51}$Nb$_{5.64}$Cu$_{1.29}$ in  weight \% , or  Fe$_{72.89}$Si$_{16.21}$B$_{6.90}$Nb$_{3}$Cu$_{1}$ in atomic \%; we will refer only to the latter in the remainder of the article.
Simulations of precipitation were performed for new compositions Fe$_{72.89 +x}$Si$_{16.21-x}$B$_{6.90}$Nb$_{3}$Cu$_{1}$ generated by varying the content of Fe and Si by $x$ ($-3 \leq x \leq 3$). Isothermal annealing was carried out at a set of temperatures between 490~\degree C and 550~\degree C in (increments of 10 \degree C) to for up to 2 hrs holding time. We obtained significant amounts of Thermocalc data for mean radius and volume fraction of Fe$_3$Si nanocrystals, \cite{CITRINE} which serves as training set for the machine learning stage of the workflow (Figure~\ref{Paper_2_Flowchart}).

\begin{figure}[ht]
\centering
\includegraphics[width = 6 cm]{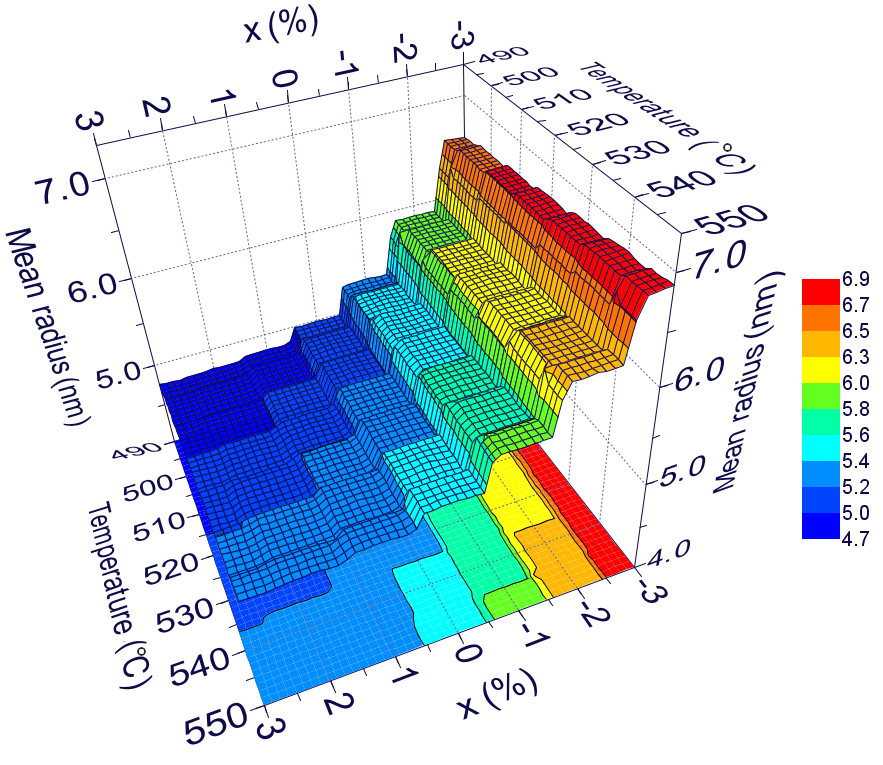}
\caption{Mean radius of Fe$_3$Si nanocrystals as a function of $x$ and  annealing temperature, for a holding time of 1~h.\cite{CITRINE}}
\label{Mean_rad_3D_Temp_X}
\end{figure}

\begin{figure}[ht]
\centering
\includegraphics[width = 6 cm]{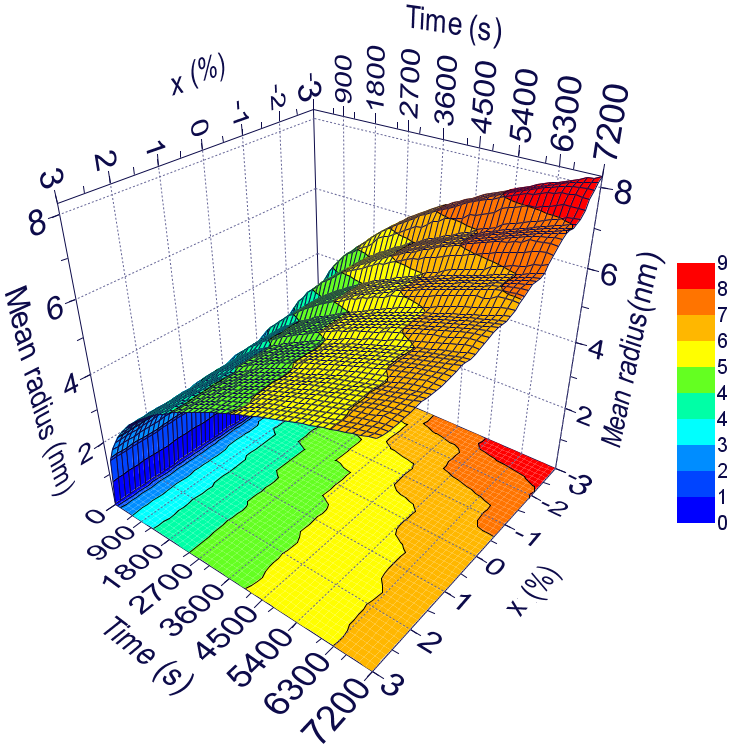}
\caption{Mean radius of Fe$_3$Si nanocrystals as a function of $x$ and  holding time, for a temperature of 500~\degree C).\cite{CITRINE} }
\label{Mean_rad_3D_Time_X}
\end{figure}

\begin{figure}[ht]
\centering
\includegraphics[width = 6 cm]{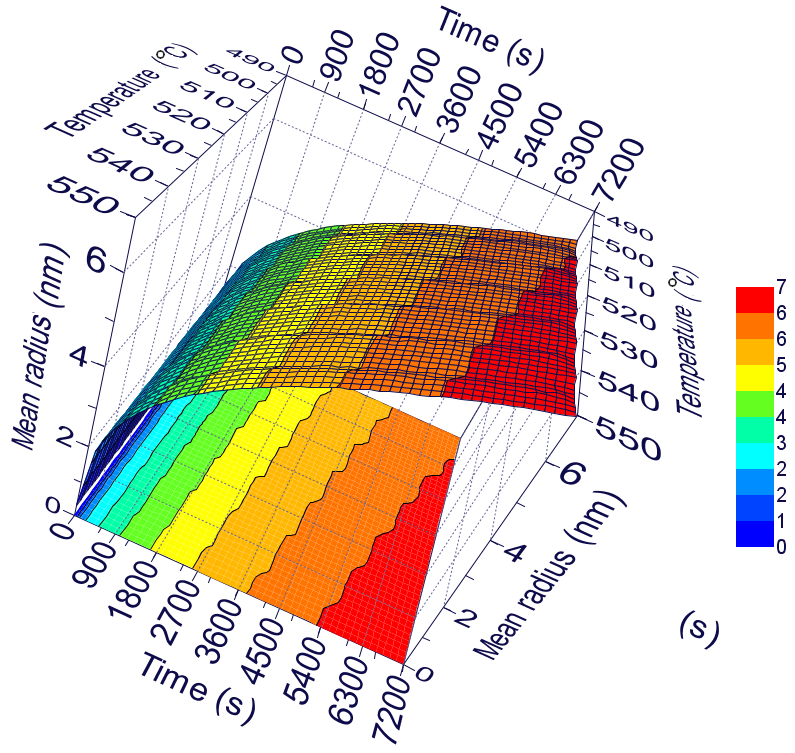}
\caption{Mean radius of Fe$_3$Si nanocrystals as a function of annealing temperature and holding time, for the nominal composition ($x=0$).\cite{CITRINE}}
\label{Mean_rad_3D_Temp_Time}
\end{figure}

\begin{figure}[ht]
\centering
\includegraphics[width = 6 cm]{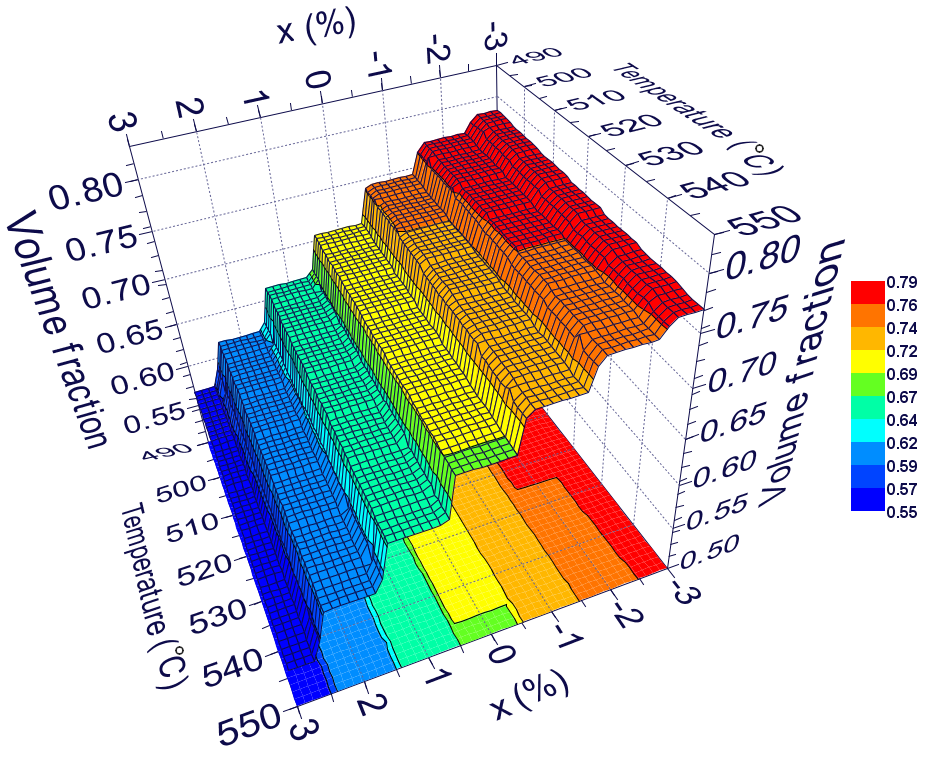}
\caption{Volume fraction of Fe$_3$Si nanocrystals as a function of annealing temperature and $x$, for 2~h holding time.\cite{CITRINE}}
\label{Volume_fraction_3D_Temp_X}
\end{figure}

\begin{figure}[ht]
\centering
\includegraphics[width = 6 cm]{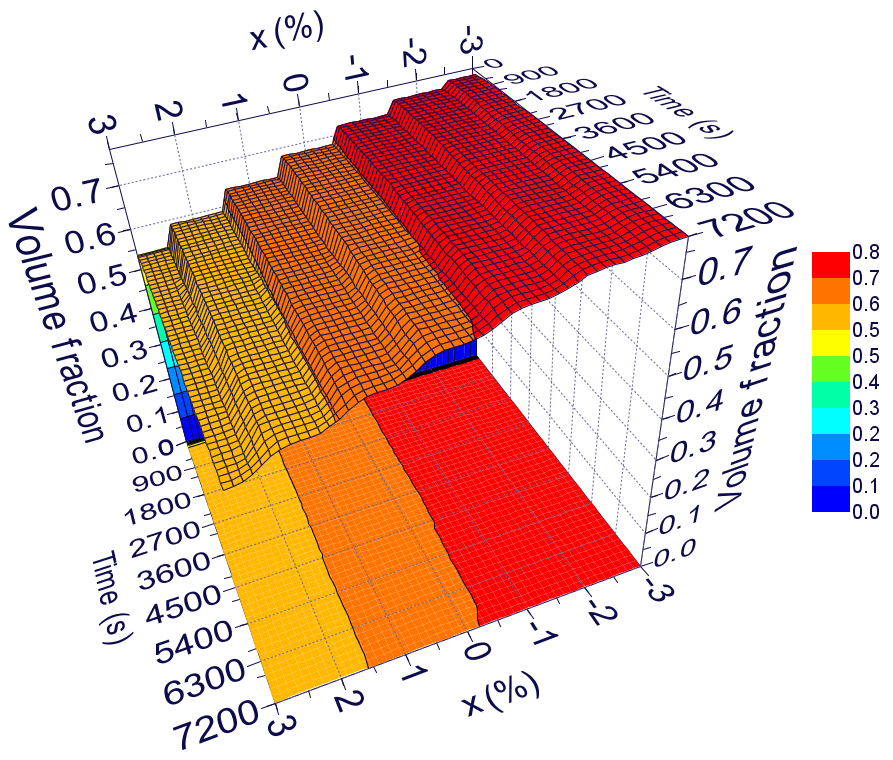}
\caption{Volume fraction of Fe$_3$Si nanocrystals as a function of holding time and $x$, at a temperature of 550~\degree C.\cite{CITRINE}}
\label{Volume_fraction_3D_Time_X}
\end{figure}

\subsection{\emph{k}-Nearest Neighbour Algorithm}\label{method_KNN}
The mean radius and volume fraction were generated in order to be used for creating response surfaces or metamodels to help
in the design of future
nanocrystalline FINEMET alloys.
We use the $k$-NN algorithm\cite{K_NN_Algorithm} as implemented in the software modeFRONTIER\cite{modeFRONTIER} to construct the metamodels.
This algorithms stores all the available information and predicts a new output (in this case, mean radius and volume fraction) based on a
measure of similarity (distance function) of the new input with the stored cases. Specifically, to predict the
new target/output that corresponds to a new input,
the straightforward approach is to compute the average of the outputs of the first $k$ nearest neighbors of the new input.
In general, the average is weighted so that some neighbors contribute more to the average than others.
In our work, we use $k = 11$ neighbors for the  metamodel describing the mean radius and volume fraction of Fe$_3$Si nanocrystals as functions of $x$,  annealing temperature, and holding time.

\subsection{Parallel coordinates chart}
A Parallel Coordinates Chart (PCC)\cite{PCC} is  a powerful tool used for
visualizing large and multivariate sets of data or results.\cite{PCC}.
It has been successfully implemented in applications such as visual and automatic data mining,
optimization, decision support, and approximations.\cite{PCC}
We find it useful to apply PCCs to materials design as well,
as a tool to visualize connections between the processing
parameters and optimized quantities.
In PCC, a number of parallel (vertical) coordinate axes
represent the $n$ dimensions of a given set of data.
Any particular data point in the $n$-dimensional space is represented by a line that connects single points
on each of the $n$ parallel coordinate axes.

We have five parallel coordinate axes in this work, three variable (input) axes, $x$, temperature, holding time, and two function (output) axes, mean radius
and volume fraction. Significant amounts of data were created, i.e.,  22,000 data sets generated from the precipitation model and 22,000 sets generated through
the metamodel for (new) randomly generated sets of variables ($x$, temperature, and time).\cite{CITRINE}
In order to properly explore the variable space, we used the Sobol algorithm\cite{SOBOL196786} for sampling the $x$, temperature, and time domains.
We analyzed all data using PCC to find the parameters that should be followed (or avoided) so as to crystallize Fe$_3$Si
nanocrystals in the desired  range for mean radius (5-7.5 nm) and volume fraction ($>$ 0.7). Another reason was to explore the possibility of decreasing holding time during isothermal annealing
without compromising on size range and volume fraction.

\section{RESULTS AND DISCUSSION}\label{results_comparison}

\subsection{Analysis of training set results and of the metamodel}

Using the data obtained from Thermocalc for few compositions and temperatures, we plot the mean radius and volume fraction as
surface meshes in which one of the variables ($x$, temperature, or time) is fixed at a certain value, while the other two are allowed to vary over their respective ranges (Figures~\ref{Mean_rad_3D_Temp_X}--\ref{Volume_fraction_3D_Time_X}).
In these mesh plots, surfaces appear stepped because they represent the training set in which $x$ and temperature each vary in seven discrete, large steps.
Even without a metamodel, these direct simulations show that one needs to avoid concentration deviations $x>0.5$\% since they would not lead to
mean radius in the desired range (Figure~\ref{Mean_rad_3D_Temp_X}).

\begin{figure}[ht]
\centering
\includegraphics[width = 7 cm]{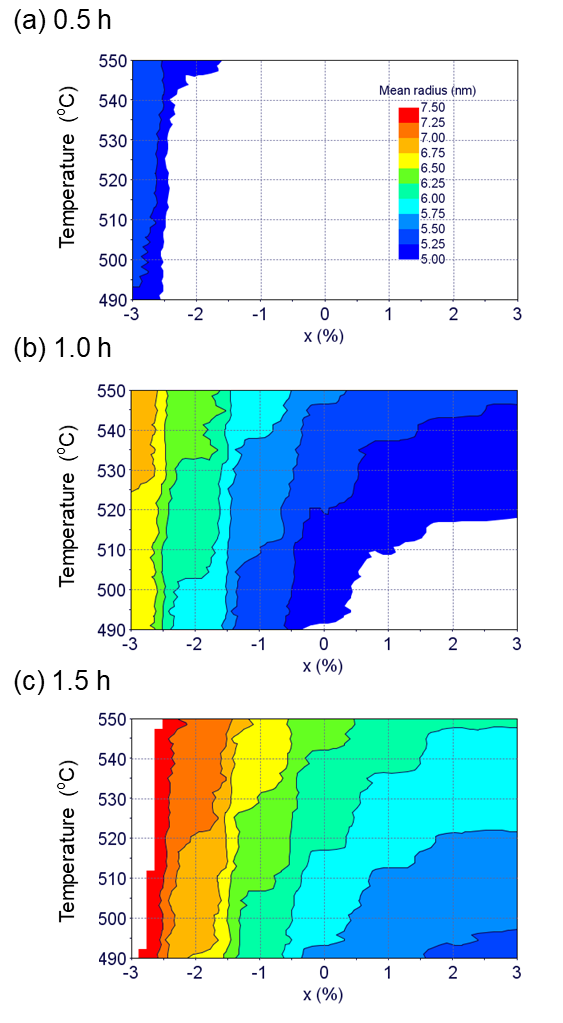}
\caption{Mean radius predicted from the metamodel after (a) 0.5 h, (b) 1.0 h, and (c) 1.5 h annealing time. Only the desired values ($5-7.5$~nm) are shown in the
contour plots.\cite{CITRINE}}
\label{Mean_radius_meta}
\end{figure}

\begin{figure}[ht]
\centering
\includegraphics[width = 7 cm]{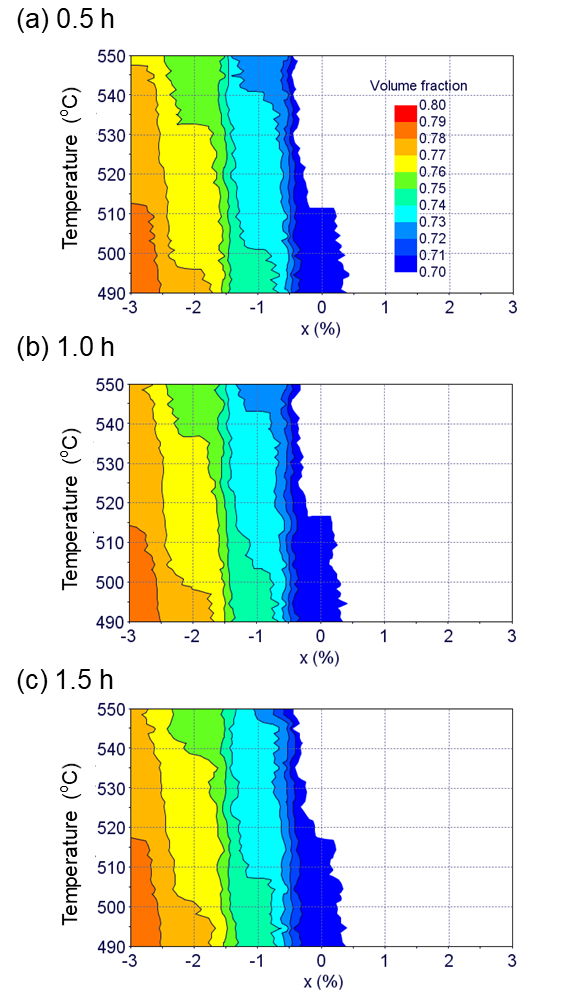}
\caption{Volume fraction of the nanocrystalline phase predicted from the metamodel after (a) 0.5 h, (b) 1.0 h, and (c) 1.5 h annealing time.
Only the desired values ($\geq 0.7$) are shown in the contour plots.\cite{CITRINE} }
\label{Volume_fraction_meta}
\end{figure}

Using the metamodels based on the $k$-NN algorithm trained on data in Figures~\ref{Mean_rad_3D_Temp_X}--\ref{Volume_fraction_3D_Time_X},
we develop contour plots for mean radius and volume fraction as functions of temperature and $x$, for three annealing times (Figures~\ref{Mean_radius_meta},~\ref{Volume_fraction_meta}).
In Figures ~\ref{Mean_radius_meta} and~\ref{Volume_fraction_meta}, we show the 22,000 data sets that have not been used in the training of the $k$-NN metamodel.
Other ways to plot the results from the metamodels so as to aid the design of the processing with ($x$, temperature, time) as variables and with two properties to
optimize (radius and volume fraction) will be shown in Sec.~\ref{result_PC_chart}.

Next, we focus on assessing the performance of the metamodels derived trained on the data in Figures~\ref{Mean_rad_3D_Temp_X}--\ref{Volume_fraction_3D_Time_X}).

\subsection{Predictions of the metamodel on Sobol sequences}

To assess the usefulness of the metamodel for design, we compared its predictions
for inputs that were not included in the training set, with Thermocalc results for the same new inputs.
To sample the three-variable space, we use Sobol sequences\cite{SOBOL196786} to generate 20,160 new random
data input points for the three processing parameters as follows: 140 points for $x$ between -3\% and 3\%,
12 points for temperatures between 490 and 550~\degree C, and 12 points for annealing times up to 2~h.

\begin{figure}[ht]
\centering
\includegraphics[width = 8.4 cm]{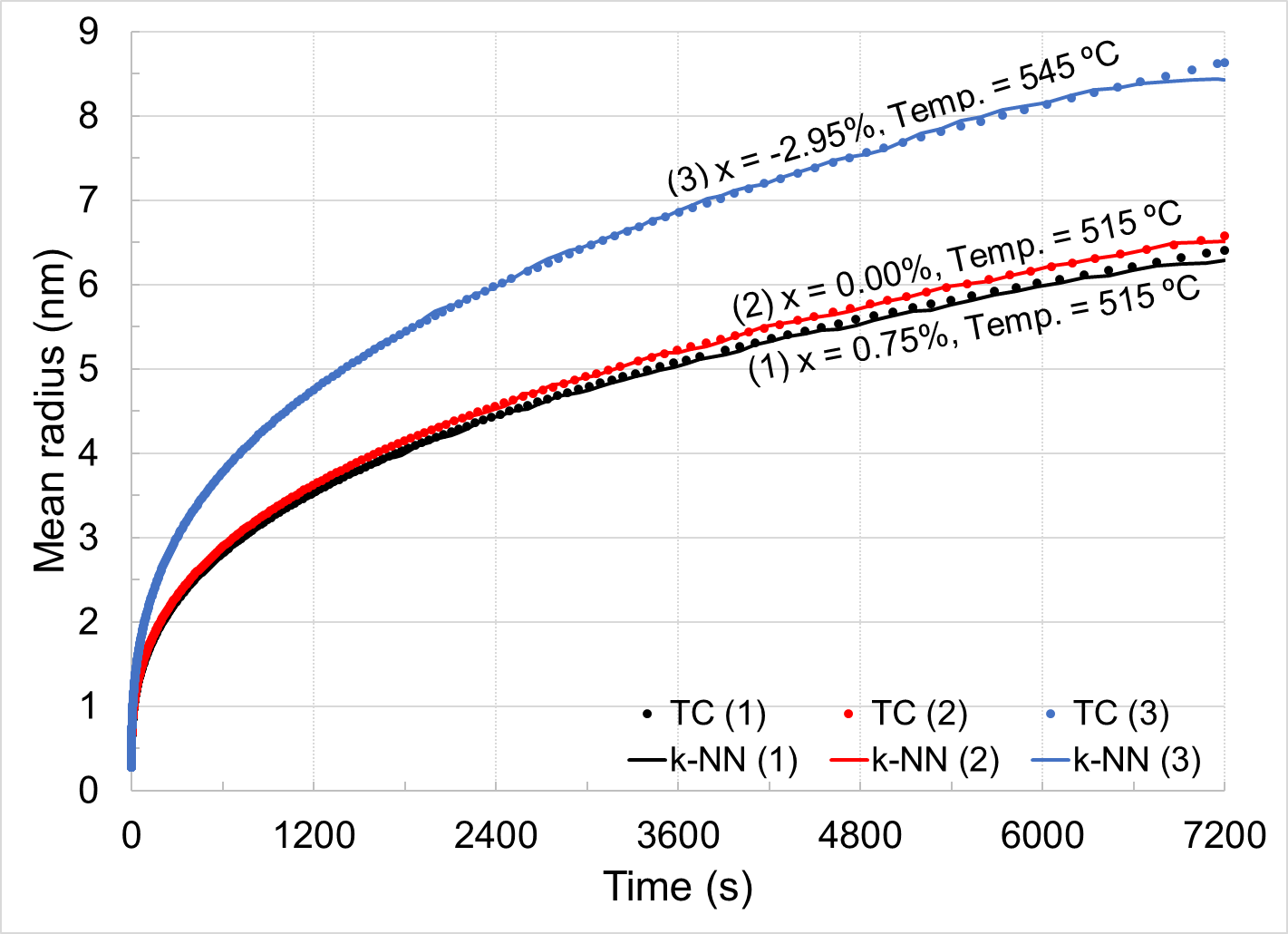}
\caption{Mean radius (Fe$_3$Si) vs. Time: Comparison between Thermocalc (TC) and metamodel (KN) prediction.\cite{CITRINE}}
\label{Comparison_TC_KNN_Mean_rad_Time}
\end{figure}

\begin{figure}[ht]
\centering
\includegraphics[width = 8.4 cm]{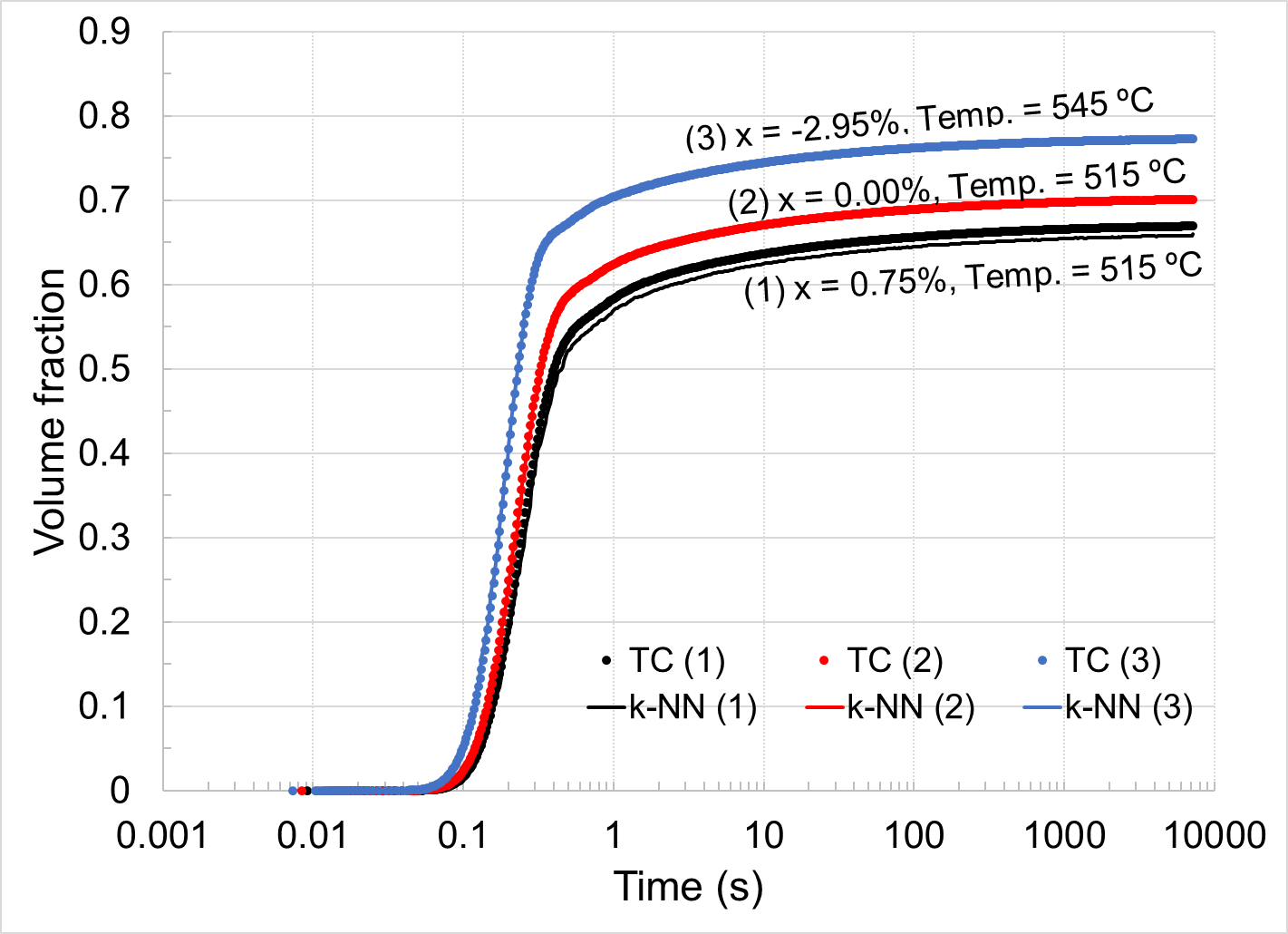}
\caption{Volume fraction(Fe$_3$Si) vs. Time: Comparison between Thermocalc (TC) and metamodel (KN) prediction.\cite{CITRINE}}
\label{Comparison_TC_KNN_Volume_fraction_Time}
\end{figure}
To illustrate comparisons between the predictions of the metamodel and direct Thermocalc results for the
new 20,160 input data based on Sobol sequences and not used in the training set, we have selected three pairs of processing
parameters, ($x$ = 0.75, 515~\degree C), ($x$ = 0, 515~\degree C) and ($x$ = $-$2.95, 545~\degree C). Figure~\ref{Comparison_TC_KNN_Mean_rad_Time} and \ref{Comparison_TC_KNN_Volume_fraction_Time} show both the metamodel predictions (curves) and the Thermocalc calculations (dots) for
the mean radius of Fe$_3$Si nanocrystals and their volume fraction, respectively.
As apparent from these figures, there are only small
deviations of the predictions of the metamodel from the direct Thermocalc results for the entire annealing duration considered.
Referring to the 3D surface plots (Figures~\ref{Mean_rad_3D_Temp_X}--\ref{Volume_fraction_3D_Time_X}), we mentioned that their stepped appearance was due to the
sparseness of the training sets for temperature and composition deviation $x$. Despite the sparseness of the training data,
we note that the developed response surfaces (metamodels) are able to capture closely the trends observed during nucleation and growth of Fe$_3$Si nanocrystals for both  mean radius (Figure~\ref{Comparison_TC_KNN_Mean_rad_Time})  and volume fraction (Figure~\ref{Comparison_TC_KNN_Volume_fraction_Time}), even though the machine learning algorithm was not exposed to the physical principles of nucleation and growth: this illustrates the predictive power of the machine learning approach in this case.
Next, we will show quantitative assessments of the accuracy of the model trained on different subsets of the original 24,000 datapoint training set.

\begin{figure}[ht]
\centering
\includegraphics[width = 8 cm]{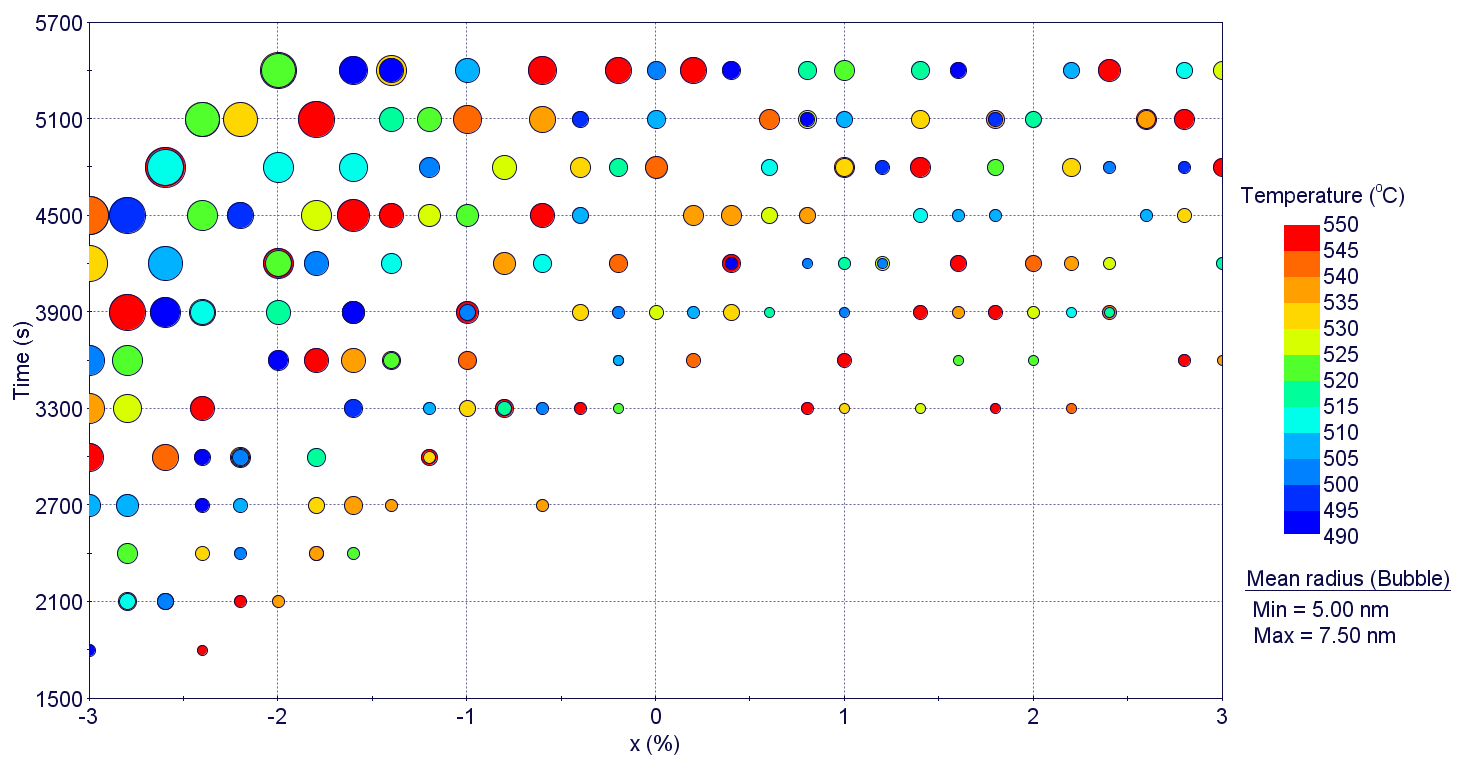}
\caption{Mean radius (circles) for different values of $x$, annealing time, and temperature. For clarity, only
a subset of 300 data sets is shown.\cite{CITRINE}}
\label{Bub_Meanrad}
\end{figure}

\begin{figure}[ht]
\centering
\includegraphics[width = 8 cm]{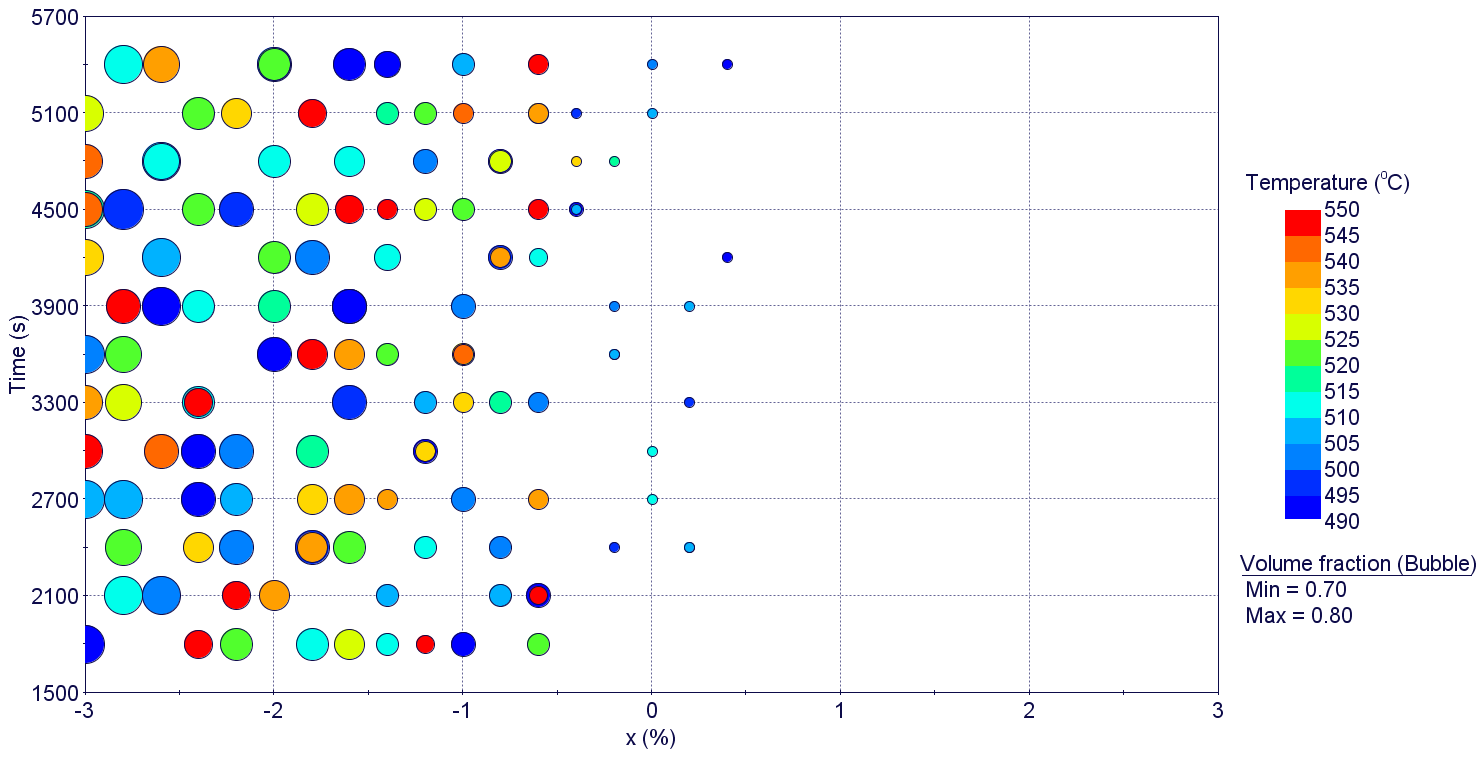}
\caption{Volume fraction for different values of $x$, annealing time, and temperature. To avoid overlaps, only
a subset of 300 data sets is shown.\cite{CITRINE}}
\label{Bub_Volfr}
\end{figure}

\subsection{Assessment of the prediction accuracy for the metamodel on different training sets}\label{subsec_test_meta}

In order to assess quantitatively the accuracy of the predictions of the metamodel, we have divided the initial 24,000 data sets into 90\%-10\% partitions (splits),
in which 90\% of the set (21,600 datasets) are used for training, and the remaining (unseen) 2,400 points are used for testing the accuracy.
We have created 20 such disjointed, random splits $(j=1,2,...,20)$, each of which containing a set of unseen inputs
${\textbf{x}_{i_j}}$, where the index $i_j=1,2,..., 2,400$ counts the testing data in split $j$ and the vector $\textbf{x}$ denotes the triplet ($x$, temperature, time).
Since both the mean radius $R$ and the volume fraction $f$ vary from near zero to a finite quantity during nucleation and
growth, we calculate the percent relative error (rather than the absolute error),
\begin{equation}
\epsilon ^{(j)} _{R,i_j} = 100\frac{|R_{\text {pred}} (\textbf{x}_{i_j}) - R_{\text {act}} (\textbf{x}_{i_j})|}{R_{\text {act}} (\textbf{x}_{i_j})},
\end{equation}
where the superscript $(j)$ indicates the split index, and the subscripts ``pred" and ``act" refer to the prediction
of the metamodel trained on the training set $j$ and the actual value computed by Thermocalc, respectively.
Similarly, the error for the volume fraction $f$ is computed as
\begin{equation}
\epsilon ^{(j)} _{f,i_j} = 100\frac{|f_{\text {pred}} (\textbf{x}_{i_j}) - f_{\text {act}} (\textbf{x}_{i_j})|}{f_{\text {act}} (\textbf{x}_{i_j})}. \label{eq:fij}
\end{equation}

For each split $j=1,2,..., 20$, we determine the minimum, maximum, average, and 95 percentile error and tabulate the results (Tables~\ref{tab_Error_mean_radius} and \ref{tab_Error_vol_fraction}). Furthemore, we compute the average and standard deviations of each of
these errors across  all the splits, and list these in the last two rows of Table~\ref{tab_Error_mean_radius} and \ref{tab_Error_vol_fraction};
for example, $\epsilon_R$ and $\sigma_R$ for the ``avrg." column in Table~\ref{tab_Error_mean_radius} are defined, respectively,
as the average and standard deviation of the average errors corresponding to the 20 splits.

We note that, overall, the accuracy of the model is very good since it yield errors almost always are within less than
1\% from the Thermocalc-computed mean radius and volume fraction
(exact values of the different types of error measures are listed in
Tables~\ref{tab_Error_mean_radius} and \ref{tab_Error_vol_fraction}). The results in the maximum error column of
Table~\ref{tab_Error_vol_fraction} may at first sight appear alarming, since they
show maximum relative errors between 8 and 13\%. Upon closer inspection of the data,
we have determined that the points responsible for these types of large (maximum) relative
errors correspond to annealing times of the order of seconds: for such small timeframes, the volume fractions
are very close to zero so any deviation from the (small) Thermocalc values $f_{\text {act}} (\textbf{x}_{i_j}) $
could trigger a large  relative error via Eq.~({\ref{eq:fij}}).
Indeed, when considering the entire 2,400 datasets in any of the testing partitions $j$,
the average error is smaller than (1/3) of a percent (Table~\ref{tab_Error_vol_fraction}).
This problem does not appear for the mean radius because $R$ (Table~\ref{tab_Error_mean_radius})
increases abruptly from zero, while the volume fraction has
nearly zero derivative at the onset (\ref{Comparison_TC_KNN_Volume_fraction_Time}).

\begin{table}[ht]
  \centering
  \caption{Relative error (\%) for mean radius for the testing set of each of the
  20 different random, disjointed splits of the 24,000 datasets into 90\% training
  and 10\% testing.}
\begin{tabular}{lllll}
\hline
\hline
     $j$ & \multicolumn{4}{c}{Relative error (\%) for mean radius} \\
    \hline
			     &min. & max. & avrg. & 95 perc.  \\
\hline
     1 & 6.3E-06 & 1.843 & 0.152 & 0.498 \\
	2	& 4.8E-05 &	1.429 &	0.147 &	0.473 \\
	3	& 2.5E-06 &	1.180 &	0.149 &	0.485 \\
   	4	& 1.6E-04 &	1.935 &	0.147 &	0.471 \\
	5	& 6.0E-06 & 1.187 &	0.145 &	0.472 \\
	6  & 9.3E-05 &	1.440 &	0.147 &	0.479 \\
	7  & 6.8E-05 &	1.214 &	0.146 &	0.475 \\
	8  & 1.9E-05 &	1.130 &	0.153 &	0.517 \\
	9  & 1.0E-06 &	1.078 & 0.142 & 0.475 \\
	10  & 7.0E-05 & 1.592 & 0.150 & 0.484 \\
	11  & 5.9E-05 & 3.138 & 0.154 & 0.498 \\
	12  & 6.0E-06 & 1.052 & 0.145 & 0.483 \\
	13  & 9.6E-05 & 1.649 & 0.152 & 0.509 \\
	14  & 6.0E-05 & 0.925 & 0.149 & 0.510 \\
	15  & 2.4E-04 & 1.800 & 0.151 & 0.487 \\
	16  & 4.0E-06 & 1.462 & 0.152 & 0.495 \\
	17  & 7.0E-06 & 2.079 & 0.152 & 0.486 \\
	18  & 4.3E-05 & 1.289 & 0.147 & 0.493 \\
	19  & 3.9E-05 & 1.480 & 0.153 & 0.517 \\
	20  & 6.0E-06 & 1.341 & 0.146 & 0.483 \\
	 \hline			
$\epsilon_R$  & 5.2E-05 & 1.512 & 0.149 & 0.490 \\
 	\hline
 $\sigma_R $  & 6.1E-05 & 0.495 & 0.003 & 0.015 \\

     \hline
 \end{tabular} \label{tab_Error_mean_radius}%
\end{table}

\begin{table}[ht]
  \centering
  \caption{Relative error (\%) for volume fraction for the testing set of each of the
  20 different random, disjointed splits of the 24,000 datasets into 90\% training
  and 10\% testing.}
\begin{tabular}{lllll}
\hline
\hline
    $j$ & \multicolumn{4}{c}{Relative error (\%) for volume fraction} \\
    \hline
			     &min. & max. & avrg. & 95 perc.  \\
\hline

1 &	        1.2E-05	&      11.532	&	0.227	&	1.190 \\
2 &		1.5E-05	&	10.21	&	0.203	&	1.136 \\
3 &		7.4E-06	&	8.600	&	0.178	&	0.778 \\
4 &		2.0E-06	&	14.63	&	0.221	&	1.106 \\
5 &		8.4E-06	&	9.71	&	0.199	&	1.006 \\
6 &		8.5E-06	&	6.58	&	0.135	&	0.630 \\
7 &		6.9E-07	&	15.30	&	0.182	&	0.891 \\
8 &		4.1E-06	&	8.95	&	0.162	&	0.709 \\
9 &		9.9E-06	&	11.04	&	0.195	&	0.889 \\
10 &	9.7E-06	&	8.30	&	0.182	&	0.897 \\
11 &	2.3E-05	&	9.99	&	0.157	&	0.810 \\
12 &	2.3E-06	&	9.38	&	0.185	&	0.879 \\
13 &	8.6E-06	&	11.06	&	0.148	&	0.678 \\
14	&	1.8E-05	&	10.60	&	0.205	&	1.078 \\
15	&	5.0E-06	&	12.09	&	0.196	&	0.871 \\
16	&	1.3E-06	&	9.53	&	0.199	&	1.057 \\
17	&	1.0E-05	&	12.56	&	0.266	&	1.379 \\
18	&	7.4E-06	&	11.96	&	0.188	&	0.755 \\
19	&	1.2E-05	&	11.52	&	0.228	&	1.294 \\
20	&	2.0E-06	&	8.93	&	0.190	&	0.985 \\
\hline				
$\epsilon_f$	&	8.4E-06	&	10.62	&	0.192	&	0.951 \\
\hline
$\sigma_f$	&	5.9E-06	&	2.09	&	0.030	&	0.204 \\

     \hline
 \end{tabular} \label{tab_Error_vol_fraction}%
\end{table}

We have tested many other sets of input parameter sets as well, and have concluded that the metamodel developed via the \emph{k}-Nearest Neighbour algorithm is sufficiently robust that its predictions can be verified back and forth for random compositions and  processing parameters. Lastly, in order to emphasize the potential usefulness of the metamodel for design purposes, we have estimate the time taken by the metamodel to drive experimental predictions for different processing parameters.
Table~\ref{Table_Comparison} shows a comparison of time estimates for experiments (annealing estimated at 1~h per sample at a given temperature), direct CALPHAD calculations, and the
metamodel: as seen in Table~\ref{Table_Comparison}, for large numbers of datasets (44,000) the time decreases from years (experiments) to minutes  (metamodel). In practice, the experiments simply would not be performed for such large datasets and researchers would necessarily have to make inferences from far fewer sets of data. The availability of a metamodel in this case allows for the rapid identification of experimental conditions/parameters that lead to optimal mean radius and volume fraction.

\begin{table*}[ht]
  \centering
  \caption{Comparison of time estimates between experiments, CALPHAD, and machine learning approach.}
\begin{tabular}{|>{\centering\arraybackslash}p{2.5cm}|>{\centering\arraybackslash}p{3.5cm}|>{\centering\arraybackslash}p{3.5cm}|>{\centering\arraybackslash}p{3.5cm}|}
\hline
\rule[-1ex]{0pt}{2.5ex}    & \multicolumn{3}{c|}{Average Annealing time } \\
\hline
\rule[-1ex]{0pt}{2.5ex} Set of parameters & Experiments & CALPHAD (Thermocalc) & Machine Learning \\
\hline
\rule[-1ex]{0pt}{2.5ex} 1 & 1 hour & 30 Seconds & Developing metamodel: Less than 2 Minutes \\
\hline
\rule[-1ex]{0pt}{2.5ex} 44,000  & 44,000 Hours = 1,833 Days = \textbf{5 Years (Just annealing)} &  366 Hours = \textbf{15.28 Days} & Less than 2 minutes
\textbf{Total: Less than 4 minutes}.
 \\
\hline
\end{tabular}
   \label{Table_Comparison}%
\end{table*}%

\subsection{Other representations of the metamodel results for use in multivariate design space}\label{result_PC_chart}
We note that contour plots such as those in Figures~\ref{Mean_radius_meta} and \ref{Volume_fraction_meta} do not fully show all
of the information that practitioners may need. In particular,  creating
 multiple time slices could make the use of the metamodels cumbersome.
 To further pursue the idea of charting the parameter space,
we are pursuing graphical ways to
represent all the correlations between the optimized quantities and the input parameters; in other words,
we develop graphical ways to show the predictions of the metamodel so as to facilitate their use for design purposes.
One such way is to represent each optimized quantity (mean radius and volume fraction) in terms of all  three
inputs using ``bubble'' plots (Figures~\ref{Bub_Meanrad} and \ref{Bub_Volfr}),
in which the inputs are on the horizontal axis ($x$), vertical axis (time), and on a color scale (temperature).
For such representations, the mean radius or the volume fraction predicted by the metamodel are shown as circles of different radii
(Figures~\ref{Bub_Meanrad},~\ref{Bub_Volfr}).  Armed with an interactive plot (such as those created in modeFRONTIER),
a practitioner seeking to design FINEMET alloys would choose a mean radius, and then simply read out from the bubble plots (Figure~\ref{Bub_Meanrad})
various possibilities (i.e., $x$, temperature, and time) leading to that mean radius.
The value added by such plots is that they can supply multiple combinations of parameters for the
same mean radius or volume fraction, thereby providing choices for experimentalists. For example,
it can be beneficial or cost-effective to choose lower temperatures and shorter annealing times, which may be enabled
by minor composition variations.

\begin{figure*}[ht]
\centering
\includegraphics[width = 12 cm]{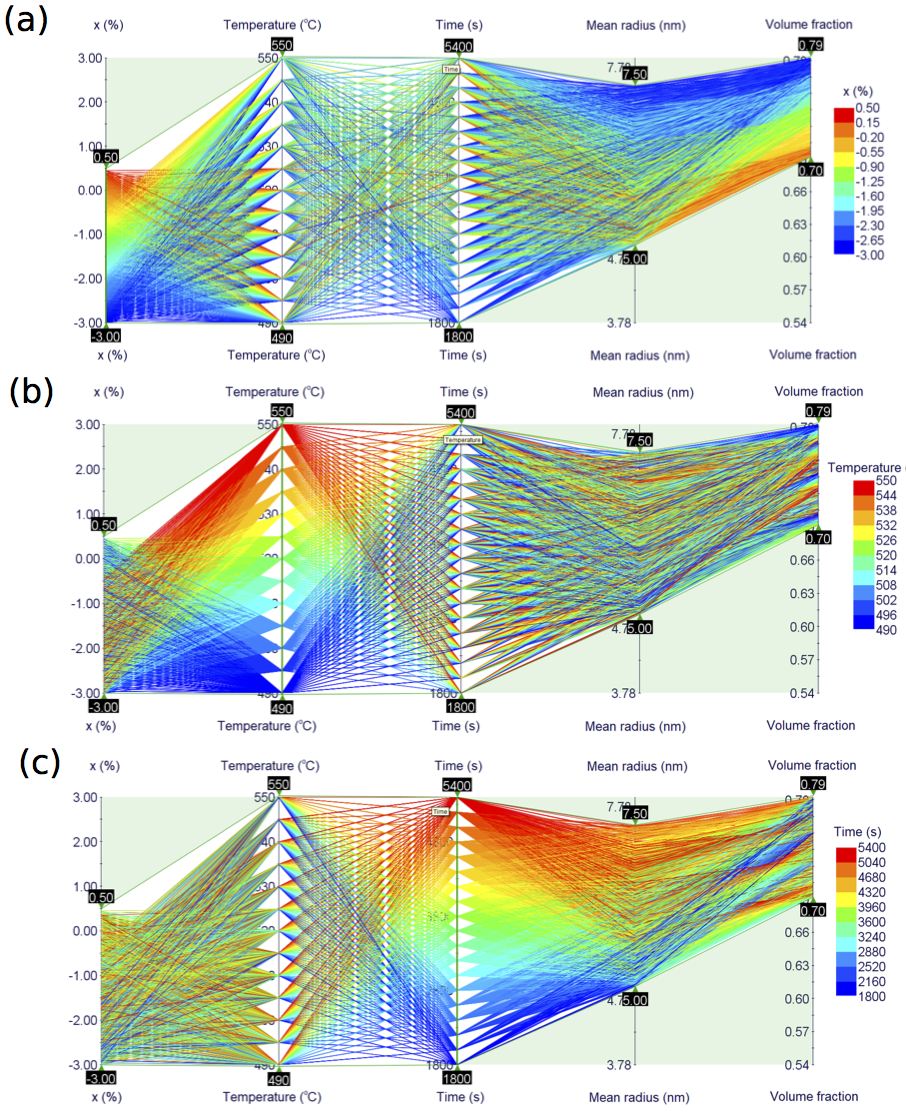}
\caption{Parallel coordinates chart for input variables (a) $x$, (b) temperature, and (c) time from 44,000 sets of data.\cite{CITRINE}}  \label{Variable-x_time_temp}
\end{figure*}

\begin{figure}[ht]
\centering
\includegraphics[width = 8.4 cm]{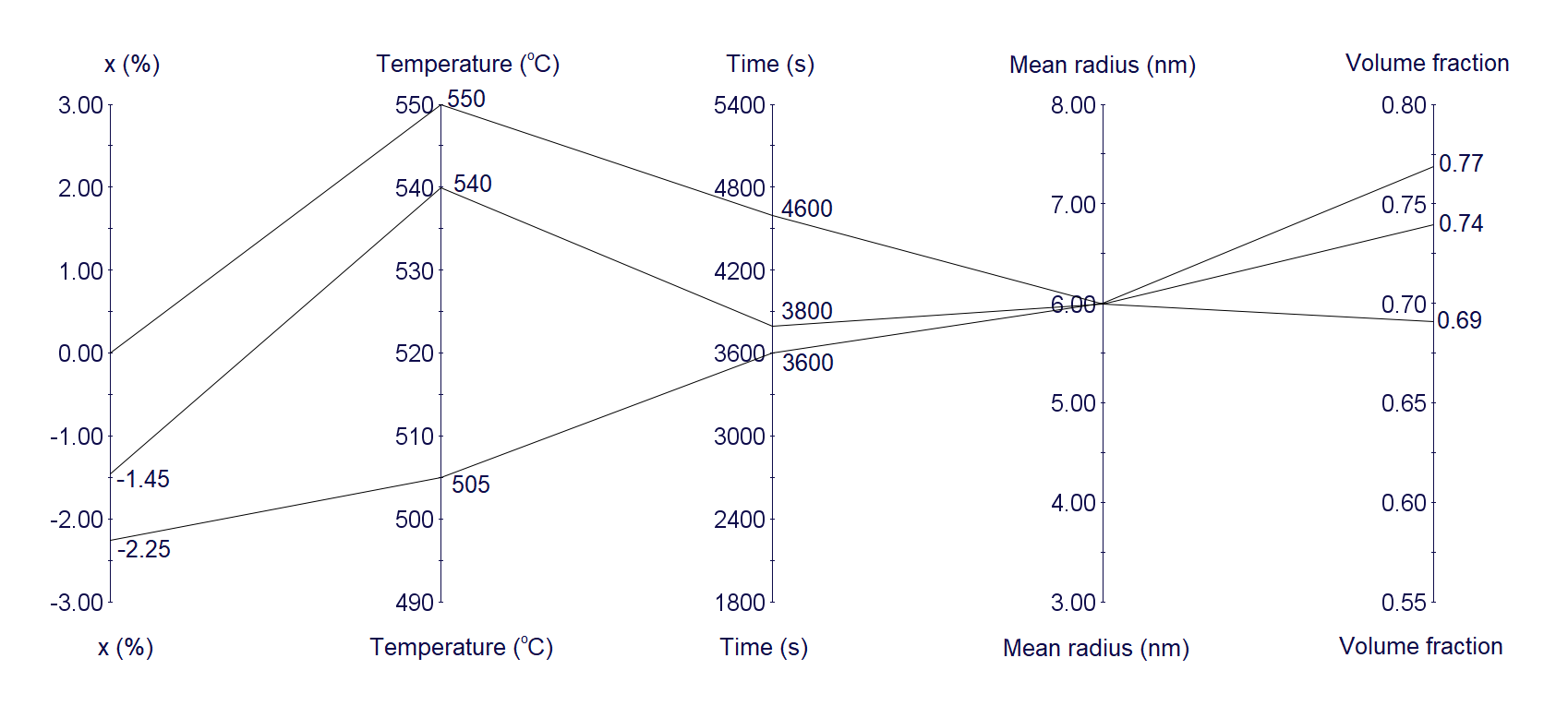}
\caption{Parallel coordinates chart for mean radius  of Fe$_3$Si nanocrystals. For clarity, this PCC shows
only three of the ($x$, temperature, time) sets that lead to one desired mean radius (6~nm).\cite{CITRINE}}  \label{onemeanrad}
\end{figure}

\begin{figure*}[ht]
\centering
\includegraphics[width = 12 cm]{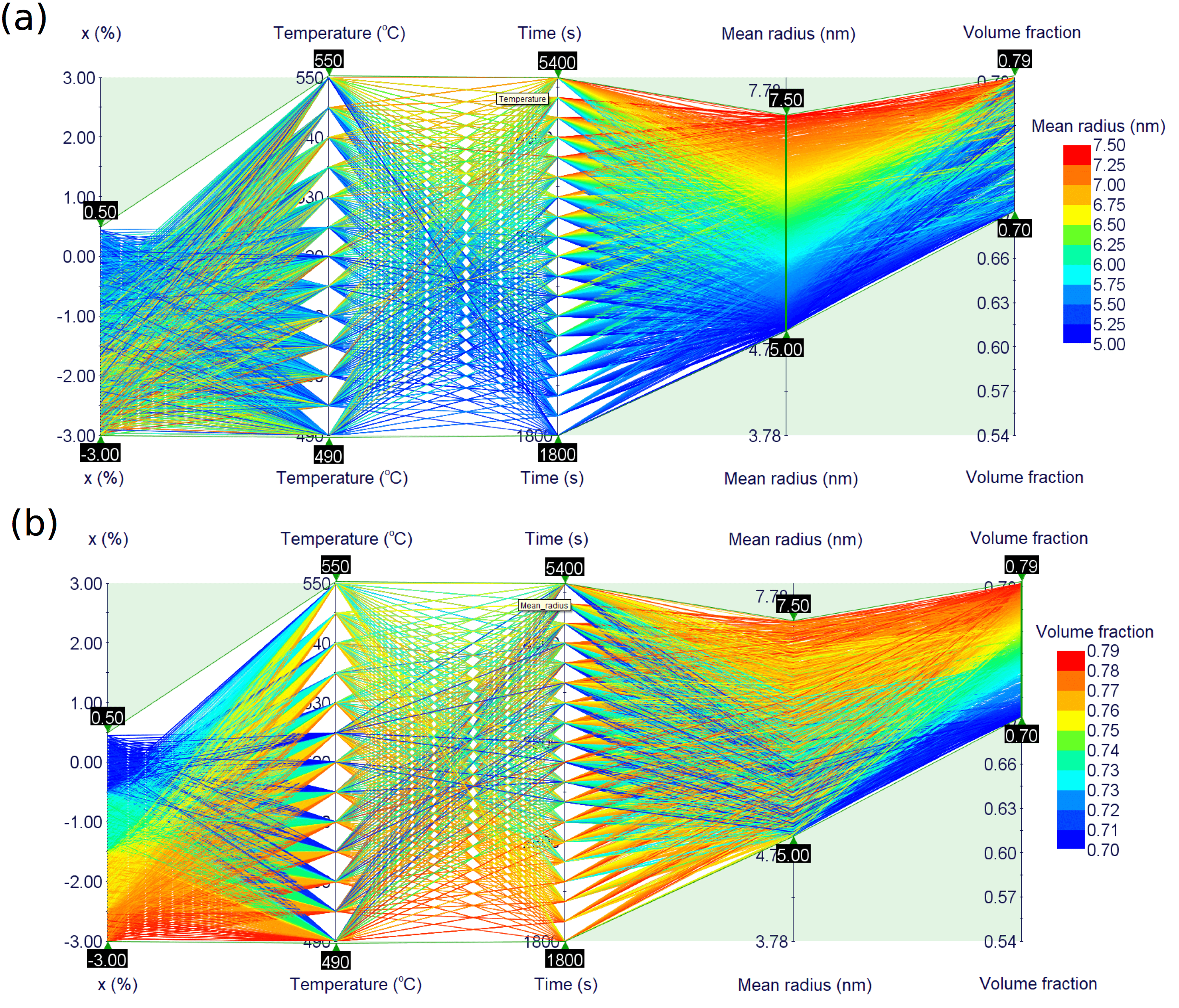}
\caption{Parallel coordinates charts obtained from 44,000 sets of data
for (a) mean radius and  (b) volume fraction (separate color legends) of Fe$_3$Si nanocrystals.\cite{CITRINE}} \label{output_m_rad_vol_fr}
\end{figure*}

While such bubble plots show each optimized quantity for all the three inputs,
 one can encounter optimization problems with more than three inputs and more than
two optimized quantities. We propose that, in general, all inputs and outputs can be shown on parallel coordinates, as mentioned in the Methods section.
PCCs\cite{PCC} are an effective way to demonstrate the relationships between desired properties (output) and the input parameters for high dimensional data,
in which the inputs and outputs stand on equal footing: any input or output quantity is shown on one of the parallel axes.
We can use PCCs to assess the effect of each separate input on the design quantities (Figure~\ref{Variable-x_time_temp}), or we can choose
design targets  and then select input parameters from the many
sets that lead to that design target (Figures~\ref{onemeanrad}, \ref{output_m_rad_vol_fr}).

Figures~\ref{Variable-x_time_temp}(a),(b), and (c) show the effect of $x$, temperature, and time, respectively on the mean radius and volume fractions.
The same 44,000  data sets \cite{CITRINE} were plotted in all three panels of Figure~\ref{Variable-x_time_temp}, with the difference that the color scale from red to blue is placed on
the different input axes for ease of separating the individual effects on mean radius and volume fraction. Consistent with the
other analysis (Figure~\ref{Mean_radius_meta}), lowering the concentration deviation $x$  (Figure~\ref{Variable-x_time_temp}(a)) increases the mean radius and the volume fraction. The effect of temperature is not so obvious from these PCCs with large amounts of data, since it is strongly coupled with the effect of annealing time.
As such, all temperatures can lead to the desired range of mean radius and volume fraction, provided that the annealing time is greater than 0.5 h (Figures~\ref{Variable-x_time_temp}(b,c)). In practice, the PCC charts are interactive: a user can simply click select a line crossing the axis
of a quantity to be optimized, and then the values for all the parallel coordinates are displayed: in particular, a large set of input triplets.
For clarity, we illustrate this point here with a very small amount of (selected) data. For clarity, in
Figure~\ref{onemeanrad} we display only three such sets of input triplets corresponding to a selected mean radius of 6 nm.
In this figure,  the nominal composition requires 550~\degree C and annealing for 1.28 h (4600 s).
Allowing the concentration to deviate from the nominal one (e.g., $x=-1.45$\% or -2.25\%) can lead to
lowering both the temperature and the annealing duration (Figure~\ref{onemeanrad}).
This is precisely the value that the metamodel brings to the design process, i.e. providing
accurate, rapidly computed input parameters that guide the
processing so as to obtain desired outputs (mean radius and volume fraction) while
often allowing for time and cost optimizations as well.
For completeness, we also present the PCCs with separate color scales for the two optimized properties (Figure~\ref{output_m_rad_vol_fr}).

\section{CONCLUSIONS}\label{conclusions}

From the results of the metamodel, there are several general
conclusions pertaining the effects of input parameters on
mean radius and volume fraction. For short anneal times ( 0.5~h), nanocrystals with the desired mean radius
 can be achieved for any temperature in the range considered, provided that $x<-2.5$\%. When annealing is sufficiently long (2~h),
then all compositions $x$ investigated can lead to mean radius in the desired range.
A volume fraction $>0.7$ can be obtained for all the investigated temperatures and for all holding
times longer than 0.5~h, provided that the composition deviation $x$ falls below the threshold of +0.5\%.

The metamodels provide not only general trends, but, more importantly, specific and multiple inputs
that lead to the desired output (mean volume and volume fraction).
These are important choices for design, which
can be carried out based completely based on the metamodel simulation results.
In particular,  the PCCs would help practitioners decide the inputs
based on their material composition  and even on economic
considerations (lower anneal temperature and/or time for large scale production).

The present results illustrate a robust approach for discovering relationships between
processing and structure/morphology in nanocrystalline alloys.
Taking a broader perspective view on these results, the combined CALPHAD-machine learning approach can be
in principle generalized for many other design situations in which structure or morphology optimization
via processing conditions is necessary.
Key factors for obtaining predictive metamodels would be to ensure all important inputs are considered, and to
judiciously select the machine
learning technique based on with the characteristics of the training data available.

\section{ACKNOWLEDGMENTS}
The authors acknowledge the financial support from the National Science Foundation through Grant. No. DMREF-1629026.

\section{Data Availability}
Data created during this work, along with the corresponding figures, has been deposited to the
repository maintained by Citrine Informatics; files are available for public access\cite{CITRINE} at https://citrination.com/datasets/154863/




\bibliographystyle{elsarticle-num}


\end{document}